\documentclass[pdftex,twocolumn,epjc3]{svjour3}          

\RequirePackage[T1]{fontenc}

\RequirePackage{fix-cm}
\smartqed  
\usepackage{graphicx}
\usepackage{comment}
\usepackage{amsmath}
\usepackage{enumitem}
\usepackage{ulem}
\usepackage{epsfig}
\usepackage{color}
\newcommand{\psixprior}{\varphi^\mathrm{prior}_x} 
\newcommand{\psixpost}{\varphi^\mathrm{post}_x} 
\newcommand{\psixaus}{\varphi^\mathrm{3B}_x} 
\newcommand{\psixno}{\varphi^\mathrm{HM}_x} 
\newcommand{\PsiTB}{\Psi^\mathrm{3B(+)}}

\def\nuc#1#2{\relax\ifmmode{}^{#1}{\protect\text{#2}}\else${}^{#1}$#2\fi}
\newcommand{\bi}{\begin{itemize}}
\newcommand{\ei}{\end{itemize}}

\newcommand{\be}{\begin{equation}}
\newcommand{\ee}{\end{equation}}
\newcommand{\vecK}{\vec{K}}

\newcommand{\vecr}{{\vec r}}

\newcommand{\vecrp}{\vec{r}\mkern2mu\vphantom{r}'}

\usepackage{xcolor}
\begin{document}

\title{The Hussein-McVoy formula for inclusive breakup revisited
}

\subtitle{A Tribute to Mahir Hussein}


\thankstext{e1}{e-mail: moro@us.es}

\author{M. G\'omez Ramos \thanksref{addr1}
        \and
         J.  G\'omez-Camacho \thanksref{addr2,addr4}
        \and
        Jin~Lei \thanksref{addr5,addr6}
        \and
        A. M.~Moro\thanksref{e1,addr2,addr3}\thanksref{}
}


\institute{Institut f\"ur Kernphysik, Technische Universit\"at Darmstadt, D-64289 Darmstadt, Germany \label{addr1} 
\and
Departamento de F\'{\i}sica At\'omica, Molecular y Nuclear, Facultad de F\'{\i}sica, Universidad de Sevilla, Apartado 1065, E-41080 Sevilla, Spain  \label{addr2}
\and 
Instituto Interuniversitario Carlos I de F\'isica Te\'orica y Computacional (iC1), Apdo.~1065, E-41080 Sevilla, Spain    \label{addr3}
\and
Centro Nacional de Aceleradores, U. Sevilla, J. Andalucía, CSIC, Avda Thomas A Edison, 7, E-41092 Sevilla, Spain  \label{addr4}
\and
School of Physics Science and Engineering, Tongji University, Shanghai 200092, China \label{addr5}
\and
Institute for Advanced Study of Tongji University, Shanghai 200092, China \label{addr6}
}

%

\date{Received: date / Accepted: date}

\maketitle

\begin{abstract}
In 1985, Hussein and McVoy [Nuc.~Phys.~A445 (1985) 124] elucidated a formula for the evaluation of the nonelastic breakup (``stripping'') contribution in  inclusive breakup reactions. The formula, based on the spectator core model, acquires a particularly simple and appealing form in the eikonal limit, to the extent that it has become the standard procedure to analyze single-nucleon knockout reactions at intermediate energies. In this contribution, a critical assessment of this formula is presented and its connection with other, noneikonal expressions discussed. Some calculations comparing the different formulae are also presented for the one-nucleon removal of $^{14}${O}+$^{9}$Be reaction at several incident energies. 

\keywords{Breakup reactions \and Inclusive breakup \and Hussein-McVoy model}
 \PACS{PACS code1 \and PACS code2 \and more}
\end{abstract}

\section{Introduction}
\label{sec:intro}
Breakup reactions have been extensively used to extract  nuclear structure information  (binding energies, spectroscopic factors, electric response to the continuum, etc) and have also permitted to improve our understanding of the dynamics of reactions among composite systems. 
When the projectile dissociates into two fragments,  the process can be described as an effective three-body problem, which can be schematically represented as 
$a+A \rightarrow b+ x+ A$,  where $a$ represents the projectile which eventually dissociates into $b+x$. Even in this simplified three-body picture, the theoretical description of the process is not  straightforward due to the presence of three particles in the final state. 

In some applications, one is interested in the inclusive process in which only one of the fragments (say, $b$) is measured experimentally, that we represent schematically as $A(a,b)X$. These inclusive cross sections are needed, for example, in the application of the surrogate method \cite{Esc12} and in spectroscopic studies by means of intermediate-energy knockout reactions \cite{Han03,Tos01,Tos14}. 

The evaluation of inclusive breakup reactions poses a challenging theoretical problem because many processes can in principle contribute to the $b$ singles cross section. When the two fragments $b$ and $x$ ``survive'' and the target remains in its ground state, the process is referred to as {\it elastic breakup} (denoted EBU hereafter), also called {\it diffraction dissociation}.

The remaining part of the inclusive breakup cross section, that we denote globally as {\it nonelastic breakup} (NEB),  includes those processes in  which the $x$ particle interacts nonelastically with the target nucleus. This involves, for example, the transfer of $x$ to a bound state of the residual system $B=A+x$, the fusion of $x$ forming a compound nucleus (incomplete fusion) or simply the target excitation by $x$. If $x$ is a composite system, it also includes any process in which the latter is broken or excited in any way. The explicit evaluation of all these processes is not possible in general so several authors proposed closed-form formulae which avoid the sum over the final states. Interestingly, all these formulae 
display a common structure, given by
\begin{equation}
\label{eq:neb}
\left . \frac{d^2\sigma}{dE_b d\Omega_b} \right |_\mathrm{NEB} = -\frac{2}{\hbar v_{a}} \rho_b(E_b)
 \langle  \varphi_x  | W_x |  \varphi_x  \rangle   ,
\end{equation}
where  $\rho_b(E_b)=k_b \mu_{b} /[(2\pi)^3\hbar^2]$ is the density of states (with $\mu_b$ the reduced mass of $b+B$ and $k_b$ their relative wave number),  $W_x$ is the imaginary part of the optical potential $U_x$, which describes $x+A$ elastic scattering. Expression (\ref{eq:neb}) offers a intuitively appealing interpretation of nonelastic breakup. As particle $b$ is scattered, fragments $x$ and $A$ interact with each other. In the original Hamiltonian, the interaction between $x$ and $A$ will be represented by a real operator depending on the internal degrees of freedom of the target nucleus. After application of a Feshbach reduction, this interaction is replaced by the complex potential $U_{xA}
$, describing $x+A$ elastic scattering and whose imaginary part accounts for nonelastic breakup events. 
Indeed, the difficulty in this interpretation is to provide a proper description of the state $|\varphi_x \rangle$ (the $x$-channel wave function hereafter), which should describe the relative motion  of $x$ and $A$, compatible with the incoming boundary conditions, and with the fact that fragment $b$ will be finally detected with a given momentum $\vec k_b$.

One of the first of these expressions 
was due to Hussein and McVoy \cite{HM85}, given explicitly in the next section. In the same work, they also derived an approximate  expression 
obtained by treating the distorted waves appearing in (\ref{eq:neb})  in the Glauber (also referred to as eikonal) approximation. Since this approximation is valid at high energies, the HM formula so obtained is expected to be accurate also at high energies. In fact, this eikonal formula is the key tool to evaluate the NEB part of the inclusive cross section in nucleon removal knockout reactions  used to study spectroscopy of nucleon hole-states. In these experiments, measured cross sections are compared with theoretical calculations for EBU and NEB, with the latter being evaluated with the Glauber version of the HM formula. These theoretical cross sections are commonly evaluated assuming single-particle wavefunctions for the removed nucleon and they are later multiplied by the required spectroscopic factors derived, for example, from shell-model calculations. Then, the ratio $R_s = \sigma^\mathrm{exp}/\sigma^\mathrm{theo}$ is computed. Typically, one obtains $R_s < 1$, which has been interpreted as an effect of
additional correlations not present in small-scale shell-model calculations,  presumably leading to a
larger fragmentation of single-particle strengths (and a subsequent reduction of spectroscopic
factors). Moreover, these studies have found a systematic dependence of this ratio on the separation energy of the removed nucleon, with $R_s$ becoming smaller and smaller as the separation
energy becomes larger \cite{Tos14}. Some authors have interpreted this result as an indication of additional correlations (coming from tensor and short-range components of the
nucleon-nucleon interaction). However, this interpretation has been recently put into question
by other authors, because this trend is apparently not observed in other reactions, such as transfer \cite{Fla13}
and $(p, pN)$ reactions \cite{Ata18,Kaw18,Gom18}.

Clearly, the conclusions strongly rely on the validity of the formulae used in the evaluation of the inclusive cross sections. Whereas there is a general consensus on the evaluation of the EBU cross sections, with different approaches leading to consistent results (as, for instance,  the distorted-wave Born  approximation (DWBA) \cite{Bau83}, the continuum-discretized coupled-channels  (CDCC) method  \cite{Aus87} and a variety of semiclassical approaches  \cite{Typ94,Esb96,Kid94,Cap04}), the reliability of the NEB calculations has been more controversial. One of the criticisms concern the validity of the Glauber HM formula at the energies commonly used in these experiments (several tens of MeV per nucleon). 

In this paper, we revisit the HM formula and its Glauber limit, and discuss its connection with other inclusive breakup formulae proposed by other authors. We present also preliminary numerical calculations comparing some of these formulae, which show that the Glauber limit of the HM  formula is a reasonable approximation at the energies at which the nucleon-knockout reactions have been measured. 

\begin{figure}
  \includegraphics[width=0.7\columnwidth]{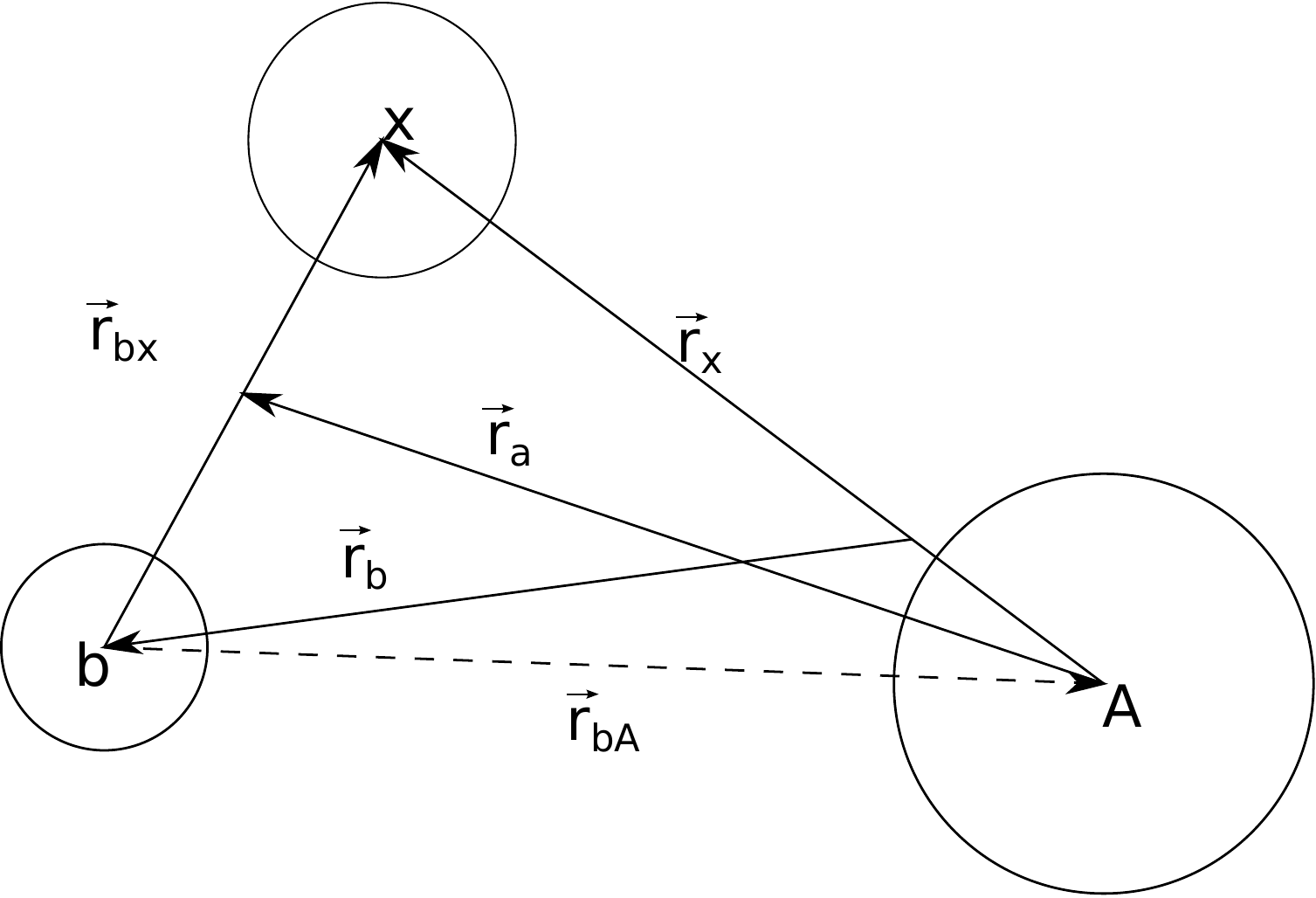}
\caption{Relevant coordinates for the three-body models described in the text.}  
\label{fig:coord}       
\end{figure}

\section{Review of inclusive breakup formulae}
\subsection{The Hussein-McVoy (HM) formula }
\label{sec:HM}
One of the first attempts to provide a closed-form formula for the inclusive breakup cross section was due to Hussein and McVoy in their seminal 1985 paper \cite{HM85}. The HM  derivation makes use of the spectator assumption for the detected fragment $b$, which means that this fragment does not participate directly in the breakup, so that the breakup is produced by any nonelastic scattering of the participant $x$ with the target. By summing over all $x+A$ states which leave $A$ in an excited state, they arrived to 
the following formula for the double differential cross section for NEB: 
\begin{equation}
\label{eq:HM}
\left . \frac{d^2\sigma}{dE_b d\Omega_b} \right |^\mathrm{HM}_\mathrm{NEB} = -\frac{2}{\hbar v_{a}} \rho_b(E_b)
 \langle  \psixno  | W_x |  \psixno  \rangle   ,
\end{equation}
where $\psixno$ is defined by the  so-called non-orthogonality  overlap, which is given by:
\begin{equation}
\label{eq:psixno}
\psixno(\vec{r}_x) = \langle \vecr_x | \psixno \rangle =   \langle \vecr_x \chi^{(-)}_b |  \chi_a^{(+)} \phi_a  \rangle .
\end{equation}
Here,  $|\phi_a \rangle $ is the projectile  ground 
state, $|\chi^{(+)}_{a} \rangle $ and  $|\chi^{(-)}_{b} \rangle$  are distorted scattering states for the $a+A$ and $b+B$ systems, respectively. $| \vec r_x \rangle $ is a state with given separation of fragment and target. The relevant coordinates are shown in Fig.~\ref{fig:coord}.

\subsection{The Eikonal Hussein-McVoy formula (EHM)}
\label{sec:EHM}
Hussein and McVoy obtained further insight on their formula (\ref{eq:HM}) by treating  the distorted waves $\chi^{(+)}_{a}(\vec{r}_a)$ and  $\chi^{(-)}_{b}(\vec{r}_b)$ in the Glauber (also known as  {\it eikonal}) approximation. The Glauber approximation to an elastic-scattering distorted wave is:
\be
\label{eq:Glauber}
\chi_{\vec  k}^{(+)}({\vec r})=\mathrm{e}^{i \vec{k} \cdot \vec{r}} \exp \left[+i \int_{-\infty}^{z} \Delta k\left(z^{\prime}, b\right) \mathrm{d} z^{\prime}\right]
\ee
where the incident momentum $\vec{k}$ points along the positive $z$-axis, and $\vec{b}$ is the component of $\vec{r}$ perpendicular to $z$. The exponent in the second factor is 
\begin{equation}
\Delta k\left(z^{\prime}, b\right) \equiv-\frac{k}{2 E} U(z, b) ,
\end{equation}
which, when integrated along the entire trajectory, gives the optical phase shift
\begin{equation}
2 \delta(b)=\int_{-\infty}^{\infty} \Delta k\left(z^{\prime}, b\right) \mathrm{d} z^{\prime}=2 \int_{0}^{\infty} \Delta k\left(z^{\prime}, b\right) \mathrm{d} z^{\prime} ,
\end{equation}
that is related to the partial-wave optical S-matrix, i.e., $S(b)=\exp[{2 i \delta(b)}]$.

A key step in the HM derivation is the choice of the $U_{a}$ potential, distorting the incident wave, which is taken as the sum of the corresponding fragment-target potentials:
\begin{equation}
\label{eq:UaA}
U_{a}=U_{bA} + U_{xA} .
\end{equation}
As we shall see, this is a particularly fortunate choice, which takes into account breakup effects in the entrance channel. Usual DWBA approaches would use a distorting potential depending only on the $a-A$ coordinate, which does not induce breakup. Indeed, they could get away with this complicated distorting potential because the eikonal approximations were assumed:
the $x$ and $b$ fragments move with the same average velocity as the projectile and hence their momenta are given by 
\begin{equation}
\label{eq:kxb}
\vec{k}_{x}=\left(m_{x} / m_{\mathrm{a}}\right) \vec{k}_{\mathrm{a}}, \quad \vec{k}_{b}=\left(m_{b} / m_{a}\right) \vec{k}_{a} .
\end{equation}

With the particular choice (\ref{eq:UaA}) and the assumption (\ref{eq:kxb}) one obtains the following result 
\begin{equation}
\begin{aligned}
\varphi^\mathrm{EHM}_x(\vecr_x) = & 
\int \mathrm{d}^{3} \vecr_{b}
\chi_{b}^{(-)*}(\vecr_b)   \chi_{a}^{(+)}(\vecr_a) \phi_{a}(\vecr_{bx})\\
=& \mathrm{e}^{i \vec{k}_{x} \cdot \vecr_{x}} \exp \left[i \int_{-\infty}^{z_{x}} \Delta k_{x}
\left(z^{\prime}, b_{x}\right) \mathrm{d} z^{\prime}\right] \\
& \times \int \mathrm{d}^{3} \vecr_{b} \mathrm{e}^{i \vec{q} \cdot \vecr_{b}} S_{bA}\left(b_{b}\right) \phi_{\mathrm{a}}\left(\vecr_{bx}\right)
\end{aligned}
\end{equation}
with $\vec{q}=\vec{k}_b - \vec{k}'_b$, the average momentum transferred in $b-A$ elastic scattering.  

The last factor is then conveniently expressed as
\be
\int \mathrm{d}^{3} \vecr_{b} \mathrm{e}^{i \vec{q} \cdot \vecr_{b}} S_{bA}\left(b_{b}\right) \phi_{\mathrm{a}}\left(\vecr_{bx}\right) \equiv 
\mathrm{e}^{i q_{\parallel} z_{x}} \tilde{\phi}_{a,b} (\vec{q}, \vec{b}_{x} ) .
\ee 

Replacing this result into (\ref{eq:HM}) one obtains for the double differential cross section: 
\begin{align}
\label{eq:EHM}
\left . \frac{d^2\sigma}{dE_b d\Omega_b} \right |^\mathrm{EHM}_\mathrm{NEB}  & = \frac{2}{\hbar v_{a}} \rho_b(E_b)
\frac{E_{x}}{k_{x}}  \nonumber \\ 
& \times \int \mathrm{d}^{2} \vec{b}_{x} \left| \tilde{\phi}_{a,b}(\vec{q},\vec{b}_{x}) \right|^{2}
\left[ 1-\left|S_{xA} (b_{x} ) \right|^{2} \right] .
\end{align} 
It should be noticed that the NEB depends only on the asymptotic properties, this is, the $S$ matrices, of the interaction of $b$ and $x$ with the target. There is no  sensitivity on the wavefunctions in the interaction region. This is a result of the eikonal approximation, plus the particular choice of the distorted interaction, which included the imaginary potential $W_{xA}$ which ultimately generates the NEB.

In many applications, one is interested in the total yield of fragment $b$, which is obtained upon integration of the previous formula over the angular and energy variables, resulting:
\begin{align}
\label{eq:EHM}
\sigma_\mathrm{NEB}^\mathrm{EHM}
= & \frac{2}{v_{\mathrm{a}}}(2 \pi)^{3} \frac{E_{x}}{\hbar k_{x}} 
 \int \mathrm{d}^{3} \vec{r}_{b} \mathrm{d}^{3} \vec{r}_{x}     \left|\phi_{a}\left(\vec{r}_{bx}\right)\right|^{2} \nonumber \\
\times & \left|S_{bA}(b_{b}) \right|^{2} \left[1-\left|S_{xA}\left(b_{x}\right)\right|^{2}\right] .
\end{align}

This equation has an appealing and intuitive form: the integrand contains the product of the probabilities for the core being elastically scattered by the target, $|S_{bA} (b_b) | ^2$, times the probability of the valence particle being  absorbed, $(1- |S_{xA}(b_x)|)^2$. These probabilities are weighted by the projectile wave function squared, and integrated over all possible impact parameters. Because of the Glauber approximation, Eq.~(\ref{eq:EHM})  is expected to be accurate at high energies (above $\sim$100~MeV per nucleon). In fact, this formula has been extensively employed 
in the analysis of intermediate-energy knockout reactions (see e.g.~\cite{Han03,Tos01,Tos14} and references therein) mostly aimed at obtaining spectroscopic information of nucleon hole states.

\subsection{The three-body (3B) model of Austern {\it et al.}} \label{sec:3b}
In \cite{Aus87}, Austern and collaborators derived a three-body formula for the inclusive breakup cross section using as starting point the post-form representation of the exact transition amplitude
%
\begin{align}
\label{eq:tpost}
\frac{d^2\sigma}{d\Omega_b E_b }&  = \frac{2 \pi}{\hbar v_a} \rho(E_b) \sum_{c} |\langle \chi^{(-)}_{b} \Psi^{c,(-)}_{xA} |V_\mathrm{post}| \Psi^{(+)}  \rangle |^2  \nonumber \\ 
 &  \times \delta(E-E_b-E^c) ,
\end{align}
where $V_\mathrm{post} \equiv V_{bx} + U_{bA}-U_{bB}$ is the post-form transition operator,  $\Psi^{(+)}$ is the system wavefunction with the incident wave in the $a+A$ channel,  and $\Psi^{c,(-)}_{xA}$ are the eigenstates of the $x+A$ system, with $c=0$ denoting the $x$ and $A$ ground states. Thus, for $c=0$ this expression gives the EBU part, whereas the terms $c \neq 0$ give the NEB contribution.   

In this model, excitations of the target are not explicitly considered (although they are effectively taken into account by means of the optical potentials $U_{xA}$ and $U_{bA}$). Thus, the total wavefunction is given by
\begin{equation}
| \Psi^{(+)} \rangle \approx |\PsiTB \phi^{0}_A \rangle , 
\end{equation}
where $|\phi^{0}_A \rangle $ is the target ground-state 
and 
$|\PsiTB \rangle $ is the solution of the three-body equation:
\begin{equation}
 [ \hat{T}_{aA} + \hat{T}_{bx}+ V_{bx} + U_{bA} + U_{xA} - E  ]  |\PsiTB \rangle =0 .
\end{equation}

Using the Feshbach projection formalism and the optical model reduction they obtain a  closed-form expression for the inclusive breakup cross section. The latter can be further split into its EBU and NEB components \cite{Kas82}. The  EBU is given by
\begin{align}
\left . \frac{d^2\sigma}{dE_b d\Omega_b} \right |^\mathrm{3B}_\mathrm{EBU} 
&= \frac{2 \pi}{\hbar v_{a}} \rho(E_{b}) \int {d\Omega_x} \, \rho(E_x)  
 \nonumber \\
&\times \left|\left\langle\chi_{b}^{(-)} \chi_{x}^{(-)}\left(\mathbf{k}_{x}\right)\left|V_\mathrm{post}\right| \PsiTB \right \rangle \right|^{2} 
\end{align}
where $\rho(E_x)$ is the density of states of particle $x$, whereas the NEB part is given by. 
\begin{equation}
\label{eq:3b}
\left . \frac{d^2\sigma}{dE_b d\Omega_b} \right |^\mathrm{3B}_\mathrm{NEB} = 
-\frac{2}{\hbar v_{a}} \rho_b(E_b)   \langle \psixaus | W_x | \psixaus \rangle .
\end{equation}
 The three-body $x$-channel wavefunction  $\psixaus(\vecr_x)$   
 is obtained by solving the inhomogeneous equation
\begin{equation}
\label{eq:phix_3B}
(E^+_x - K_x - {U}_x)  \psixaus(\vecr_x) = \langle \vecr_x \chi_b^{(-)}| V_\mathrm{post}| \PsiTB \rangle
\end{equation}
where $E_x=E-E_b$.  This equation  can also be written in integral form as
 \begin{align}
\label{eq:phix_3B_G}
 \psixaus(\vecr_x) & =  G^{(+)}_x \langle \vecr_x \, \chi_b^{(-)}| V_\mathrm{post}| \PsiTB  \rangle \nonumber \\
 &= \int d^3 \vecrp_x \langle\vecr_x | G^{(+)}_x | {\vecrp_x} \rangle  \langle \vecrp_x \, \chi_b^{(-)}| V_\mathrm{post}|  \PsiTB  \rangle 
\end{align}
where $G^{(+)}_x= (E^+_x - K_x - {U}_x-i \epsilon)^{-1}$ is the optical model Green's function of particle $x$.

Austern {\it et al.} derived also an interesting alternative expression for $\psixaus(\vecr_x)$, namely,
\be
\label{eq:chiPsi}
\psixaus (\vecr_x) = \langle \vecr_x \chi_b^{(-)}  |  \PsiTB \rangle .
\ee

It is worth noting that the formulae (\ref{eq:chiPsi}) and (\ref{eq:phix_3B_G}) are formally equivalent and, as such, they must provide identical results. Although (\ref{eq:chiPsi}) may seem simpler, in practice, it requires that the three-body wavefunction $\PsiTB$ be accurate in the full configuration space, since there is no natural cutoff in the integration variable $\vecr_b$. By contrast, in Eqs.~(\ref{eq:phix_3B}) and (\ref{eq:phix_3B_G}),  the presence of the $V_\mathrm{post}$ operator will tend to emphasize small $b-x$ separations and hence one requires only an approximate three-body wavefunction accurate within that range.   This can be achieved, for instance, expanding $\PsiTB$  in terms of $b-x$ eigenstates, as done in the CDCC method \cite{Aus87}, or in terms of Weinberg states \cite{Pan13}. The implementation of the method with CDCC wavefunctions is numerically challenging and the first calculation of this kind was only recently reported \cite{Jin19}. 


\subsection{The Ichimura, Austern, Vincent (IAV) formula} \label{sec:iav}
Ichimura, Austern and Vincent \cite{IAV85} proposed a simpler DWBA version of the 3B formula above. In DWBA, the exact wavefunction $\Psi^{(+)}$  is approximated by the factorized form:
\begin{equation}
|\Psi^{(+)}\rangle \approx |\chi_a^{(+)}  \phi_a   \phi^{0}_A \rangle .
\end{equation}
 With this approximation, the NEB component of the $b$ singles cross section becomes
 \begin{equation}
\label{eq:iav}
\left . \frac{d^2\sigma}{dE_b d\Omega_b} \right |^\mathrm{IAV,post}_\mathrm{NEB} = 
-\frac{2}{\hbar v_{a}} \rho_b(E_b)   \langle \psixpost | W_x | \psixpost \rangle   ,
\end{equation}
that is formally identical to (\ref{eq:3b})  but with the $x$-channel wavefunction given now by 
 \begin{equation}
\label{eq:phix_post}
 \psixpost(\vecr_x) =  G^{(+)}_x \langle \vecr_x \, \chi_b^{(-)}| V_\mathrm{post}| \chi_a^{(+)} \phi_a  \rangle .
\end{equation}

The  IAV model has been recently revisited by several groups \cite{Car16,Jin15,Pot15} and its accuracy assessed  against experimental data with  considerable success \cite{Jin17,Pot17}.  

\subsection{The Udagawa, Tamura (UT) formula} \label{sec:ut}
Udagawa and Tamura \cite{Uda81} derived a formula similar to that of IAV, but making use of the prior
form DWBA. Their final result can be again expressed in the form (\ref{eq:neb}),   
\begin{equation}
\label{eq:UT1}
\left . \frac{d^2\sigma}{dE_b d\Omega_b} \right |^\mathrm{UT}_\mathrm{NEB} = 
-\frac{2}{\hbar v_{i}} \rho_b(E_b) 
 \langle \psixprior | W_x | \psixprior \rangle ,
\end{equation}
where $ \psixprior$ is the solution of a $x-A$ inhomogeneous equation similar to Eq.~(\ref{eq:phix_post}) but replacing in the source term the post-form transition operator, $V_\mathrm{post}$, by its prior form counterpart, $V_\mathrm{prior} \equiv U_{xA} + U_{bA}-U_{a}$, i.e.:
\begin{equation}
\label{phix_prior}
(E^+_x - K_x - {U}_x)  \psixprior(\vecr_x) =  \langle \vecr_x \, \chi_b^{(-)}| V_\mathrm{prior}| \chi^{(+)}_{a} \phi_a \rangle.
\end{equation}

\section{Relation among theories}

In this section we discuss the connection between the formalisms outlined in the previous section, with emphasis in the HM formulae. Some of these relations have already been discussed in previous works, most notably in the comprehensive work of Ichimura \cite{Ich90}. We focus here on the HM formulae (\ref{eq:HM}) and (\ref{eq:EHM}) due their relevance in the analysis and interpretation of knockout studies. 

The HM formula (\ref{eq:HM}) can be readily obtained starting from Austern's identity (\ref{eq:chiPsi}), and making the replacement $| \PsiTB  \rangle \approx |\chi_a^{(+)} \phi_a \rangle $.  It is also enlightening to see the connection of the HM formula (\ref{eq:HM})  with the IAV model. For that, one needs to transform first the IAV formula, Eq.~(\ref{eq:iav}) into its prior form. This can be done using the following relation due to Li, Udagawa and Tamura \cite{Li84}:
\begin{equation}
\label{eq:LUT}
\psixpost(\vec r_x)  = \psixprior(\vec r_x)  +  \psixno(\vec r_x)  .
\end{equation}
Replacing (\ref{eq:LUT}) into Eq.~(\ref{eq:iav}) one gets 
\begin{align}
\left . \frac{d^2\sigma}{dE_b d\Omega_b} \right |^\mathrm{IAV,prior}_\mathrm{NEB} & = 
\left . \frac{d^2\sigma}{dE_b d\Omega_b} \right |^\mathrm{UT}_\mathrm{NEB} + 
\left . \frac{d^2\sigma}{dE_b d\Omega_b} \right |^\mathrm{HM}_\mathrm{NEB}  \nonumber \\
&+ \left . \frac{d^2\sigma}{dE_b d\Omega_b} \right |^\mathrm{IN}_\mathrm{NEB} ,
\label{eq:post_prior}
\end{align}
%
with the interference (IN) term
\begin{equation}
\label{eq:int}
\left . \frac{d^2\sigma}{dE_b d\Omega_b} \right |^\mathrm{IN}_\mathrm{NEB} = 
- \frac{4}{\hbar v_a} \rho_b(E_b) 
 \mathrm{Re}  \langle \psixprior | W_{xA} | \psixno \rangle .
\end{equation}

Equation (\ref{eq:post_prior}) represents the post-prior equivalence of the NEB cross sections in the IAV model, with the RHS corresponding to the prior-form expression of this model. The first term is just the NEB formula proposed by Udagawa and Tamura \cite{Uda81}, which is formally analogous to the IAV post-form formula (\ref{eq:iav}), but with the $x$-channel wave function given by $\psixprior(\vecr_x)$. The second term in (\ref{eq:post_prior}) is the  HM formula of Eq.~(\ref{eq:HM}) and the last term arises due the interference between the UT and HM terms. 

The HM formula [Eqs.~(\ref{eq:HM}) and (\ref{eq:psixno})], is then recovered from (\ref{eq:post_prior}) by neglecting altogether the UT term. This result suggests that this HM formula is an incomplete NEB theory, since the UT term can be very large, or even dominant \cite{Jin15b}%
\footnote{This conclusion was in fact shared by M.~Hussein himself who, at least in recent years, was aware of the incompleteness of the HM formula \cite{Hussein}. }. 

The situation for the eikonal HM formula, Eq.~(\ref{eq:EHM}), is qualitatively different. In this case, we cannot start from the DWBA formula (\ref{eq:post_prior}), since the latter assumes that the auxiliary potential potential $U_a$ is a function of the $a-A$ relative coordinate. This is not the case of the HM choice, Eq.~(\ref{eq:UaA}). If $U_a$ is allowed to be a function of both $\vecr_a$ and $\vecr_{bx}$ coordinates  (\ref{eq:LUT}) becomes:
\begin{align}
\label{eq:LUT_2}
\psixpost & = G_x \langle \vecr_x \, \chi_b^{(-)}| V_\mathrm{prior}| \psi^{3B(+)}_{xb} \rangle 
 + \langle \vecr_x \chi^{(-)}_b |  \psi^{3B(+)}_{xb}   \rangle .
\end{align}

But, for the HM choice of the $U_a$ potential, Eq.~(\ref{eq:UaA}), $V_\mathrm{prior}$ vanishes identically, and one recovers Austern's formula, Eq.~(\ref{eq:chiPsi}). 
This shows that the EHM approximation (\ref{eq:EHM}) incorporates three-body effects which go beyond its non-eikonal form (\ref{eq:HM}). In fact, 
Eq.~(\ref{eq:EHM}) represents the Glauber limit of the  three-body formula by Austern {\it al.}. In this regard, one may expect the eikonal HM formula to be more accurate than its original noneikonal counterpart, whenever the Glauber approximation is justified (i.e., at high energies).

\section{Application to nucleon knockout from $^{14}$O}
In this section, we present some preliminary numerical results comparing the HM, EHM and IAV formulae. A more systematic study, including more detailed observables, such as momentum distributions, is in progress and will be presented elsewhere. 

\subsection{Practical considerations}
Some remarks on the numerical implementation of the different models are in order. The  post-form IAV formula faces the problem of the marginal convergence of the integrals appearing in the source term of Eq.~(\ref{eq:phix_post}), due to the oscillatory character of the functions appearing in the initial and final states. 
To overcome this problem, several regularization procedures  have been used in the literature.  Huby and Mines  \cite{Hub65} and Vincent \cite{Vin68} multiply the source term by an  exponential convergence factor, that damps the contribution of the  integral at large distances. Vincent and Fortune  \cite{Vin70}  use  integration in the complex plane to transform the oscillatory functions into decaying exponentials.  One may also replace the distorted waves $\chi_b^{(-)}(\vecr_b)$ by wave packets constructed by averaging these distorted waves over finite energy intervals \cite{Jin15,Jin15b}. The resulting averaged functions become square-integrable and the source term of Eq.~(\ref{eq:phix_post}) vanishes at large distances.

Notice that this problem does not arise in the prior form version of this formula because, in this case, the transition operator ($V_\mathrm{prior}$) makes the source term short-ranged. In \cite{Jin15b}, a numerical comparison between the post and prior IAV formulae was performed for the  $^{58}$Ni(d,p)X reaction and they were found to be yield almost identical results, provided a regularization procedure was applied to the post formula.

\subsection{Numerical results}
For these test calculations we consider the one-neutron and one-proton removal processes taking place in the $^{14}$O+$^{9}$Be reaction at two different incident energies, namely, 53~MeV/u and 80~MeV/u. Experimental data for this reaction were reported  in Ref.~\cite{Fla12} and analyzed in terms of the eikonal model as well as the semiclassical ``transfer the continuum'' method of Bonaccorso and Brink \cite{Bon88,Bon91}, which provides an alternative to the eikonal method for the case of neutron removal. 

In the calculations presented here, we focus on the NEB part of the cross section. We compare the DWBA IAV  [Eq.~(\ref{eq:iav})],  the HM  [Eq.~(\ref{eq:HM})] and the Eikonal  HM formulae [Eq.~(\ref{eq:EHM})]. Owing to the aforementioned convergence issues, the calculations with the IAV method were performed using its prior form formula. Intrinsic spins are ignored for simplicity. For a meaningful comparison, the same structure inputs and optical potentials are considered in all these calculations. In particular, the wavefunction of the removed nucleon is generated with a Woods-Saxon potential with parameters $R_0=3.29$~fm, $a=0.67$~fm and the depth adjusted to reproduce the proton ($S_p=4.6$~MeV) or neutron ($S_n=23.2$~MeV) separation energy, as appropriate. We used energy-independent neutron-target and core-target potentials derived from the $t \rho$ and $t \rho \rho$ approximation, respectively, assuming for the nucleon-nucleon t-matrices the parametrization of \cite{Hof78} with nucleon-nucleon cross sections from \cite{Cha90} and imaginary-to-real ratios from \cite{Ray79} and densities extracted from Hartree-Fock calculations with the SkX interaction \cite{Bro98}. For the IAV and HM calculations, the potential between projectile and target is also needed. This potential has been computed via a folding of the neutron-target and core-target potentials with the square of the wave-function between neutron and core in the projectile.  For the EHM calculation, the optical limit of the Glauber approximation was used to derive the required S-matrices from these potentials. 

\begin{figure}
  \includegraphics[width=0.8\columnwidth]{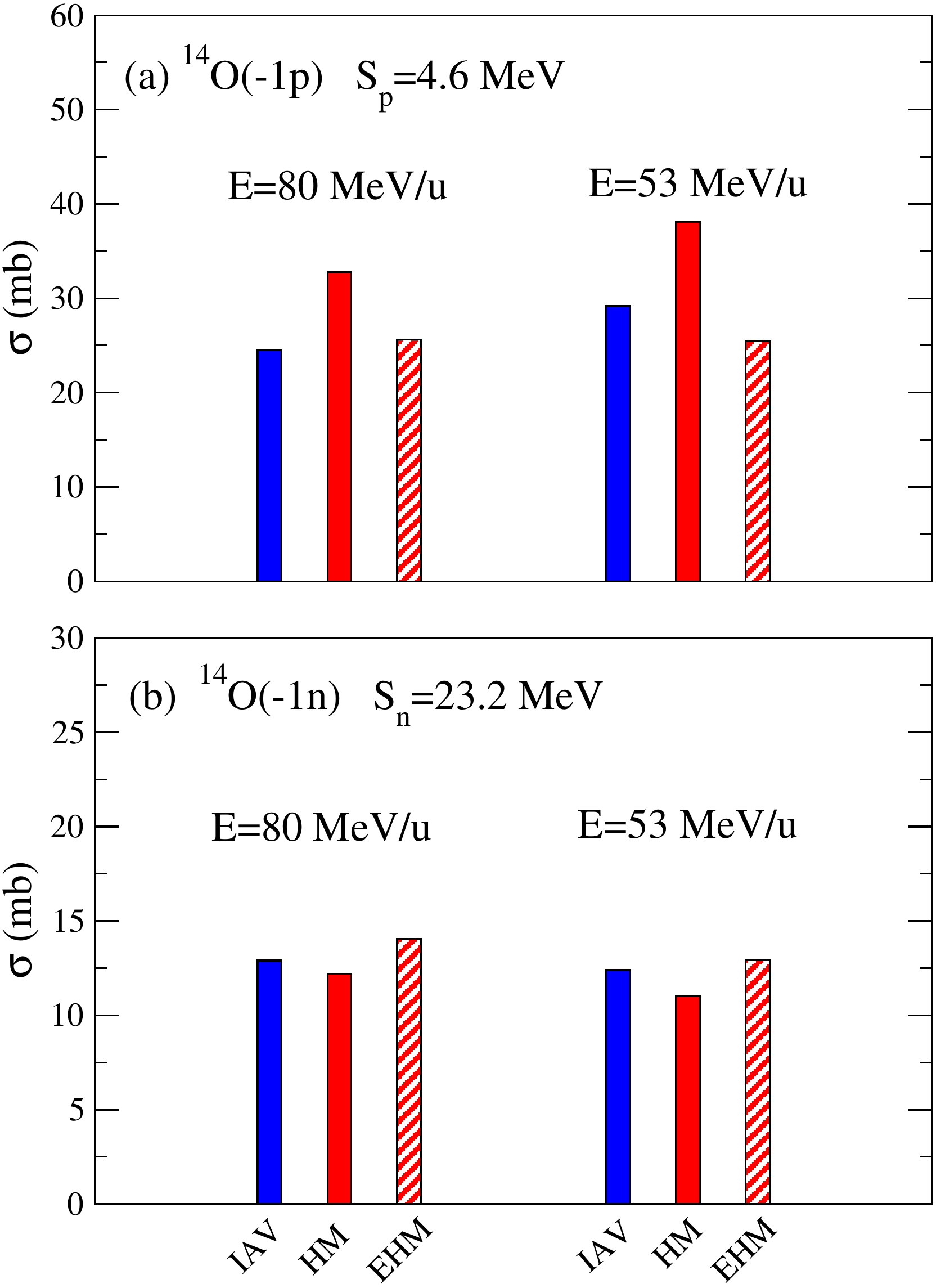}
\caption{Integrated nonelastic breakup cross sections for the $^{9}$Be($^{14}$O,$^{13}$N) (top) and $^{9}$Be($^{14}$O,$^{13}$O) (bottom) reactions at 80~MeV/u and 53~MeV/u computed with the IAV, HM and EHM formulae.}  
\label{fig:comp}       
\end{figure}

The results of these calculations are presented in Fig.~\ref{fig:comp}, in the form of bar diagrams. The upper and bottom panels correspond to the one-proton and one-neutron removal reactions.  For the neutron removal case, we find that the three methods give close results. This suggests that the nonorthogonality term, which is absent in the HM calculation, is small in the case of removal of a tightly bound nucleon. For the proton removal, the HM result tends to overestimate the IAV result. This indicates that, at lest in this case, the HM and nonorthogonality terms interfere destructively (c.f.\ last term in Eq.~(\ref{eq:post_prior})]. Remarkably, the Glauber version of the HM formula is in better  agreement with the IAV result than the original (i.e.~non-eikonal) HM   formula. As discussed in the previous section, this result can be interpreted recognizing that the  EHM can be regarded as an approximation to the full three-body IAV theory and, as such, includes effectively contributions from the three terms in  Eq.~(\ref{eq:post_prior}).  

Two main conclusions can be drawn from this  preliminary analysis. First, the non-eikonal HM formula provides an accurate approximation to the more elaborated IAV formula for deeply bound nucleons, but fails for more weakly bound nucleons. Second, the EHM formula represents a very good approximation to the IAV formula both for well bound and weakly bound nucleons and for energies as low as 50~MeV/u. This result seems to add support to the use of the EHM formula in the analysis of knockout reactions.

We note that some other approximations are implicit in the derivation of all discussed formulae, including the IAV one. For example, all these formulae are based on the spectator assumption for the core fragment. This means that the latter simply scatters elastically by the target nucleus, but does not participate in the nucleon-removal  dynamics. For example, possible rescattering effects of the struck nucleon are not taken into account. These rescattering effects, not considered by any of the presented theories, could modify the cross section for nucleon removal, whose systematics are currently under discussion \cite{Tos14}. A recent comprehensive review on the subject is given in Ref.~\cite{Aum20}.

\section{Summary and conclusions}
In this work, we have reexamined the nonelastic breakup formula devised by Hussein and McVoy (HM) and extensively employed in knockout studies. We have shown that this formula can be derived from the Ichimura, Austern, Vincent (IAV) model, recently revisited and applied by several groups, as well as from the three-body formula of Austern {\it et al.} \cite{Aus87}. We have also shown that, owing to the particular choice of the auxiliary interaction $U_{aA}$, the eikonal version of the HM formula (EHM) incorporates genuine three-body effects. These effects are also present in the more general three-body formula of Austern {\it et al.}, but in a more complicated way. 

Preliminary calculations for the one-nucleon removal in the $^{14}$O+$^{9}$Be reaction shows that the EHM formula reproduces accurately the results of the IAV model. For the removal of the strongly bound neutron ($S_n=23.2$~MeV), the noneikonal HM formula is also in good agreement with the IAV result. However, for the one-proton removal ($S_p=4.6$~MeV), the  noneikonal HM formula tends to overestimate the IAV result. It will be interesting to extend these calculations to other systems and energies to see whether these conclusions remain.

\begin{acknowledgements}
This work has been partially supported by the Spanish Ministerio de Ciencia, Innovaci\'on y Universidades under projects FIS2017-88410-P and RTI2018-098117-B-C21 and by the European Union's Horizon 2020 research and innovation program under Grant Agreement No.\ 654002. J.L. acknowledges support from the National Natural Science Foundation of China (Grants No. 12035011, No. 11975167, and No. 11535004), by the National Key R\&D Program of China (Contracts No. 2018YFA0404403 and  2016YFE0129300), and by the Fundamental Research Funds for the Central Universities (Grant No. 22120200101). M.~G.-R. acknowledges support from the Alexander von Humboldt foundation.
\end{acknowledgements}

\bibliographystyle{spphys}       
\bibliography{inclusive_epj.bib}

\end{document}